# Pressure control of conducting channels in singlewall nanotube networks

M. Monteverde
Centre de Recherches sur les Très Basses Températures, Centre Nationale de la Recherche Scientifique, 38042 Grenoble , France **and**
Laboratorio de bajas Temperaturas, Departamento de Física, Universidad de Buenos Aires, Argentina

M. Núñez-Regueiro
Centre de Recherches sur les très Basses Températures, Centre Nationales de la Recheche Scientifique, 38042 Grenoble, France

We measure electrical transport on networks of single wall nanotube ropes as a function of temperature $T$, voltage $V$ and pressure up to 22GPa. We observe Luttinger liquid (LL) behavior, a conductance $\propto T^\alpha$ and a dynamic conductance $\propto V^\alpha$. With pressure conductance increases while $\alpha$ decreases, enabling us to test the theoretical prediction for LL on the $\alpha$ dependence of the $T$ and $V$ independent coefficient of the tunneling conductance, and to obtain the high frequency cut-off of LL modes. The possible transition to a fermi liquid at $\alpha \to 0$ is unattainable, as nanotubes collapse to an insulating state at high pressures.
61.46.+w, 73.63.Fg, 62.50.+p, 75.30.Kz

Theoretical calculations have predicted that the tunnelling rate into a LL should follow the lines of the more general quantum tunnelling of double well systems with ohmic dissipation [1,2,3]. They showed that the modes of the LL should play a role analogous to the Caldeira-Leggett [4] oscillators phenomenologically used to model dissipation and that their high frequency cut-off $\omega$ could be identified with the frequency of small oscillations in each well [5,6]. In other terms, it should, first, bear the signature of the LL power law suppression of the energy $\varepsilon$ density states, $\rho(\varepsilon) \propto \varepsilon^\alpha$ (with $\alpha$ inversely proportional to the number $N$ of conducting channels) [7,8], and, second, be inversely proportional to the same power $\alpha$ of the LL modes high frequency cut-off [6]. The first assertion was verified in measurements on SWNT that yielded, in the high bias $V$ limit, a dynamical conductance $dI/dV \propto V^\alpha$, and in the low bias limit, a temperature $T$ dependence of the conductance $G = G_0 T^\alpha$ [9,10] ( with $G_0$ a constant independent of $T$ and $V$). However, the latter condition, namely $G_0 \propto \omega^{-\alpha}$ [2,3,5,6], has not been verified, because all the experiments performed on metal-isolated SWNT rope - metal devices have yielded exponents corresponding to only one Dirac cone [9,10], $\alpha \approx 0.3$. Due to the weak coupling among the tubes in the rope the tunnelling into only one SWNT dominates the measured conductance of the devices. We show here that pressure increases this coupling increasing the number of LL modes, i.e. we are able to modulate the number of Caldeira-Leggett oscillators and test the second assertion.

The samples used in this study were prepared using the electric arc discharge technique [11]. The three presented samples (0.07x0.05x0.007cc) were either measured as received (sample B) or previously transformed into a film by compressing with a watch glass (A and C). The electrical resistance measurements were performed in a sintered diamond Bridgman anvil apparatus using a pyrophillite gasket and two steatite disks as the pressure medium [12]. The Cu-Be device that locked the anvils does not allow measurements releasing the pressure and could be cycled between 4.2K and 300K in a sealed dewar.

In Fig.1 we show the temperature dependence of the conductance at different pressures for sample A. We note that in this log-log plot there is always a clear linear regime at low temperatures that, at high pressures, extends for all the temperature range. Previous reports had described a variable range hopping (VRH) behavior either of a 3-D [13] or a 2D [14,15] character that do not fit our data. In previous hydrostatic measurements, the sample is submerged in a liquid that can hinder good rope-rope contact, while we ensure contacts by sandwiching ropes and leads at high stresses between two soft steatite disks, allowing optimal lead-sample and rope-rope intra-sample contacts. The size of our samples also matters, as thinner samples (0.05x0.01x0.001cc) gave invariably 2D VRH, i.e. maximizing the size of the samples statistically ensures a percolation path through a LL network.

On Fig. 2 we plot the dynamical conductance measurements for sample A at 10GPa. In the inset we show the $dI/dV$ curves as a function of bias voltage at different temperatures. At high biases, we see that all curves coincide onto a $V^\alpha$ dependence. The LL theory imposes that the differential conductance must follow the universal scaling curve [2,3,5,6,9]

$$G \equiv dI/dV,$$

$$G(V,T) = A\left(\frac{2\pi k_B}{\hbar \omega}\right)^\alpha T^\alpha \cosh\left(\gamma \frac{eV}{2k_B T}\right) \frac{1}{|\Gamma(1+\alpha)|} \left|\Gamma\left(\frac{1+\alpha}{2} + \gamma \frac{ieV}{2\pi k_B T}\right)\right|^2 \quad (1)$$

where $\Gamma(x)$ is the gamma function, $k_B$ the Boltzmann constant, $e$ the electron charge, $A$ is a constant that, including the sample geometrical factor, and $\gamma$ is related to the inverse of the number of junctions in the sample, more precisely, the inverse of the summation of the junctions weighted by their resistances. Our data collapse completely onto the universal law. From the fit of (1) on normalized $dI/dV$ we extract an $\alpha = 0.055 \pm 0.001$ to be compared with $\alpha = 0.059 \pm 0.005$ from $G \propto T^\alpha$. The same type of accord is obtained for all pressures and samples. The fit to eq. (1) gives us additional information, in terms of the $\gamma$ parameter. As we are measuring a SWNT mat, we do not expect to be probing a single rope, but rather an interconnected network of similar ropes. We obtain a $\gamma = 0.0267 \pm 0.005$ for the shown 10GPa data, that implies that we are measuring at least 38 rope-rope junctions. This value oscillates between 37 and 40 for the different pressures, i.e. is constant with pressure within the experimental error.

The picture that we obtain from our numerical analysis is easily transposable to the typical scanning photograph of our samples (Fig. 1 of ref. 11): we have a large number of percolation paths in parallel between the electrodes, each one with a minimum of $\approx 40$ LL-LL junctions among the ropes in series. We could expect a conductivity as $T^\alpha$ for each junction, but not necessary all of them with the same $\alpha$. Calculations[16] show that if we assume a distribution of $\alpha$ near a most probable value $(\alpha_0)$ we should observe a behavior of the conductivity as $T^{\langle\alpha\rangle}$. The value of $\langle\alpha\rangle$ differs from $\alpha_0$ less than 1% for square distributions of $\alpha$ as large as 30% of $\alpha_0$ (using typically values for $\alpha_0$). Thus, our measurements yield an average $\alpha$ and in the following $\alpha \equiv \langle\alpha\rangle$.

We plot the value of $\alpha^{-1}$ as a function of pressure on Fig. 3 for samples A,B and C and observe that it is linear. Its definition [17], $\alpha_{bulk} = (1/g + g - 2)/8N$, suggests that the number of channels participating to the conduction increases linearly with pressure, if we assume that the variation of the $g$ parameter in our pressure range is negligible. Theory predicts though $\alpha \propto \sqrt{N}$ at large $N$ due to inter-tube interactions[17,18], so that the linear behaviour that we observe cannot be easily interpreted. Remarkably, the extrapolation to zero pressure is identical for the three samples. It yields a value of $\alpha(P=0) \cong 0.6$, that corresponds to twice the value of $\alpha_{bulk}$, as has been measured for two bulk contacting SWNT nanotubes[19], implying that the succession of rope-rope junctions that we are measuring in all our samples are of the bulk-bulk type.

The increase in the average number of channels with pressure can be due to an augmentation with pressure of the number of hole occupied bands, i.e. a variation of the doping $n$, holes per carbon atom, originated by impurities such as, e.g. oxygen[20]. We tentatively calculate[16] the emptying of the different bands of the typical nanotubes [21] of our sample (all tubes of diameters betwen 1.2 and 1.8nm were considered) with a constant $dn/dP$ and a $g = 0.28$, from the value of $\alpha_{bulk}(0)$. We see on Fig. 3 such a fit (dashed line) for sample C, with $dn/dP \cong 0.00075 \pm 0.00005 holes GPa^{-1}C^{-1}$. Though the fit is excellent for only one parameter, the change in the Fermi energy of $\sim 1.15 eV$ at $20 GPa$ seems exceedingly large for our pressure range (if we consider samples A or B this change would be even greater). Thus, the most probable reason for the decrease of $\alpha$ is an increase of inter-tube coupling, where the volume, including the nanotubes that furnish channels to conduction, increases with pressure, and the different numerical slopes are probably due to different rope diameter.

Although we obtain a similar linear variation of $\alpha^{-1}$ with pressure for all the samples and an identical ordinate to the origin, the numerical value of its slope is sample dependent. Different samples seem to contain different variety of ropes, and those of sample A appear to have a larger number of channels, besides being the only that presents an anomaly at 13.2GPa. This anomaly consists in a decrease of the conductance with increasing pressure (inset of Fig. 3) accompanied by the concomitant decrease of $\alpha$ shown in Fig. 3, that seem to signal a transition to a more insulating state at higher pressures. The fact that conductance drops, can be explained by a collapse of some of the conducting nanotubes, as totally squashed nanotubes are all non-metallic [22]. In fact, structural phases transitions have been detected either by

Raman or crystallographic measurements under pressure on SWNT ropes. Most reports coincide in that there is a first critical pressure region around 2GPa involving a reversible polygonalization of the walls of the nanotubes due to the increase of the pressure induced inter-tube interaction within a rope[23 24 25]. Some measurements performed above 10GPa have described another structural transitions[26], possibly implying deformations to elliptical or flattened cross-sections[27]. We observe a high pressure transition to an insulating state in only one sample, that with smaller $\alpha$, that possibly means a larger number of conducting channels, i.e. thicker ropes or larger nanotubes in this sample. We conclude that structural transitions under pressure can be very sample dependant, explaining the differences between different reports. In our particular case, thicker bundles or bundles with larger nanotubes seem to be more prone to collapse as expected[28]. The collapse is time dependent. In the inset of Fig. 4 we observe a logarithmic in time behavior for $\alpha^{-1}$, typical of relaxation in disordered systems, e.g. spin glasses[29], implying a distribution of relaxation (collapsing) times for the nanotubes. The limit $\alpha \to 0$ should correspond to a fermi liquid (FL). Though the decrease of $\alpha$ with pressure could allow us to observe the passage form the LL to a FL, the flattening transition of the SWNT leads to a semiconducting state. In any way, the conservation of LL properties up to about a score of channels would imply incoherent hopping between nanotubes[30], probably precluding such a transition.

From the conductance versus temperature data, $G(V=0,T) = G_0(P)T^{\alpha(P)}$, we can extract $G_0(P)$, which common sense would expect to increase with the number of accessible channels. The relation with the $\alpha$ parameter has been calculated and is, according to eq. (1)

$$\frac{G_0(\alpha)}{A} = \left(\frac{2\pi k_B}{\hbar\omega}\right)^\alpha \frac{1}{|\Gamma(1+\alpha)|} \left|\Gamma\left(\frac{1+\alpha}{2}\right)\right|^2 \qquad (2)$$

By fitting this expression to our data we find that $A$ is sample dependent but that $\omega$ is the same for all the samples. On Fig. 4 we see how all the samples (even the time evolution of the highest pressure of sample A) fall on the same curve, which we have chosen to plot as a function of $\alpha^{-1}$, with an unique $\hbar\omega = 2.6 \pm 0.8 eV$ that, according to Kane and Fisher [6], is the high frequency cutoff $\hbar\omega \approx E_F$ of the LL modes. Though this energy may be slightly pressure and sample dependant, the experimental error includes this variation and the obtained value agrees with the expectations.

In summary, by tuning on the appropriate inter-rope contacts, application of pressure on macroscopic samples of SWNT ropes allows the observation of the LL properties of carbon nanotubes, previously detected only in nanoscopic devices fabricated from isolated bundles. Furthermore, pressure increases the number of conducting channels, seemingly by increasing the number of nanotubes that participate in transport in each rope, producing an $\alpha^{-1}$ exponent variation that is linear in pressure. The experiment shows an alternative way to study LL in SWNT ropes without any aggressive solvating treatment that may alter the intra-rope inter-nanotube labile interactions.

We are grateful to Saïd Tahir and Patrick Bernier for providing us with the samples and to F.W.J. Hekking, and O. Buisson for a critical reading of the manuscript. M.M. is a CONICET from Argentina doctoral fellow.

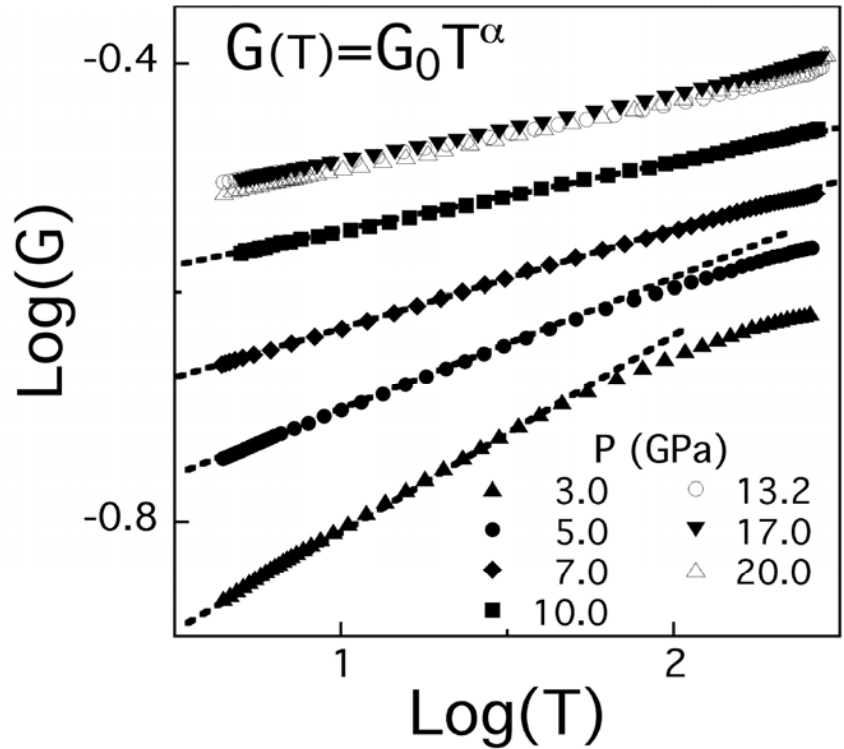

**Figure 1.** Temperature dependence of the conductance of sample A for different pressures. We observe a linear behaviour (dashed line) at low temperatures that extends to all temperature at high pressures.

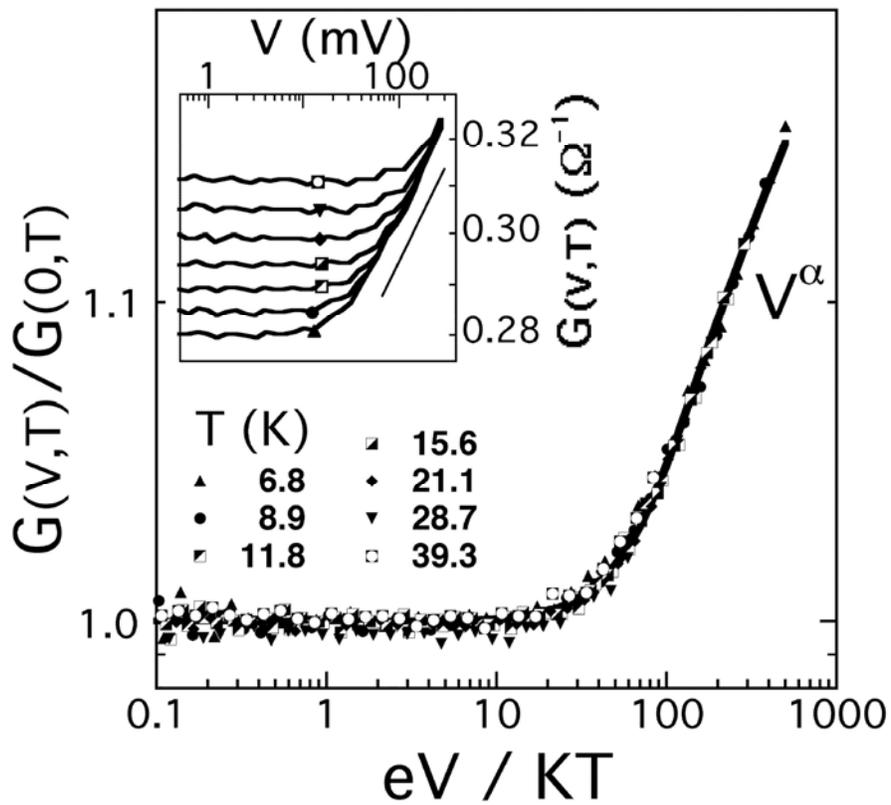

Figure 2. scaling of the dynamical conductivity of sample A at a pressure of 10GPa. We observe that the curves for all temperatures collapse to the same curve, that has a $V^\alpha$ dependence at high bias. Inset :same but as measured dynamical conductance curves at different temperatures.

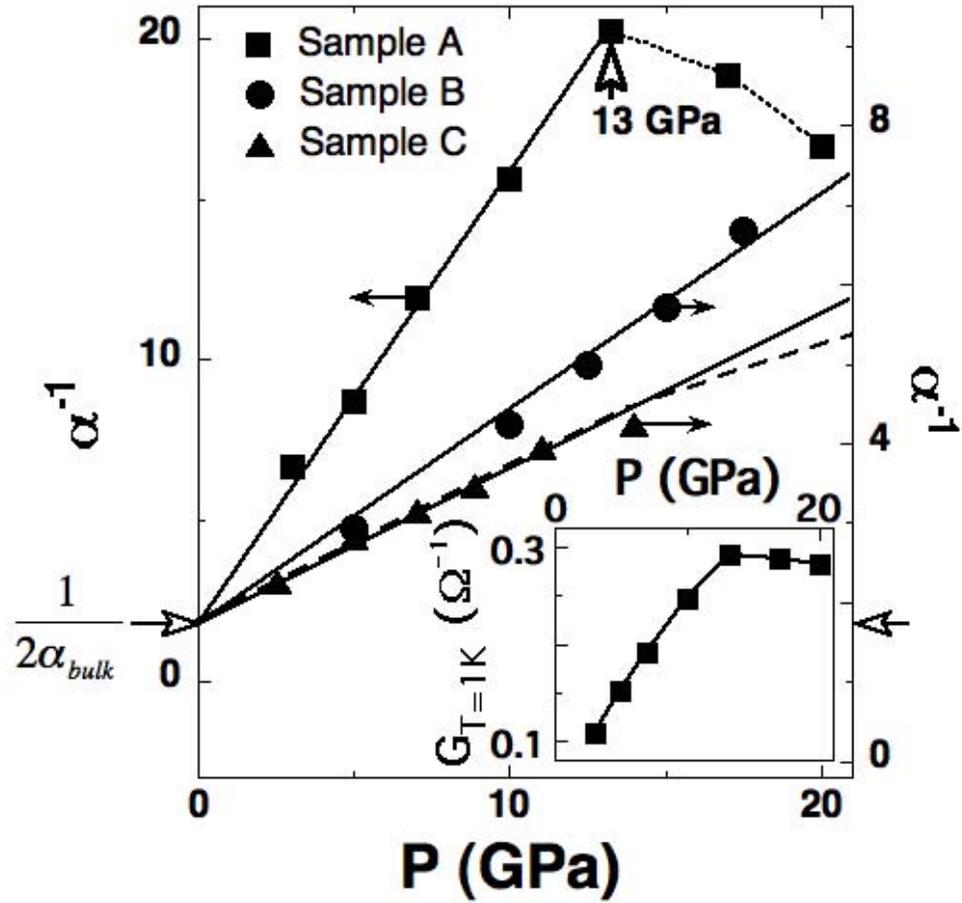

**Figure3. Pressure dependence of the $\alpha^{-1}$ parameter. We use two scales : left for the higher $\alpha^{-1}$ values sample A and right for the other, that coincide at the convergence coordinate of their dependences. We observe a linear variation for all samples that converges to the previously measured value of $\alpha^{-1}_{bulk-bulk}$ at zero pressure. The dashed fit on sample C corresponds to the one obtained with a constant $dn/dP$ under pressure (see text). Sample A shows a phase transition towards a less conducting state at 13GPa. Inset : variation of the conductance at 1K of sample A showing the phase transition at 13 GPa.**

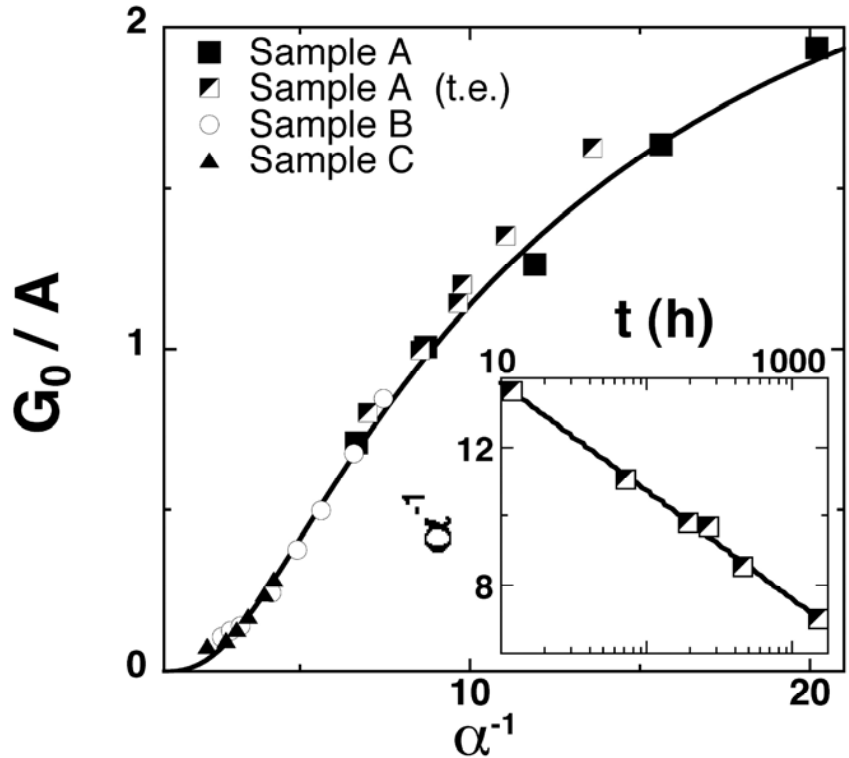

Figure 4. Relation between the value of the coefficient of the conductance temperature power law with the $\alpha^{-1}$ parameter. We observe that all samples fit to the same dependence after normalization with a geometrical factor. The universal fitting factor $\hbar\omega = 2.6 \pm 0.8 eV$ corresponds to the fermi energy of the individual nanotubes. Inset : Time evolution of the $\alpha^{-1}$ parameter of sample A at 22.1GPa (final pressure). We observe a logarithmic time dependence typical of relaxation in disordered systems.

---

[1] A.J. Leggett et al., *Rev. Mod. Phys.* **59**, 1(1987)